\pdfoutput=1
\RequirePackage{ifpdf}
\ifpdf 
\documentclass[pdftex]{sigma}
\else
\documentclass{sigma}
\fi

\DeclareMathOperator{\Tr}{Tr}

\begin{document}


\renewcommand{\thefootnote}{$\star$}

\newcommand{\arXivNumber}{1609.06439}

\renewcommand{\PaperNumber}{094}

\FirstPageHeading

\ShortArticleName{Invitation to Random Tensors}

\ArticleName{Invitation to Random Tensors\footnote{This paper is a~contribution to the Special Issue on Tensor Models, Formalism and Applications. The full collection is available at \href{http://www.emis.de/journals/SIGMA/Tensor_Models.html}{http://www.emis.de/journals/SIGMA/Tensor\_{}Models.html}}}

\Author{Razvan GURAU}

\AuthorNameForHeading{R.~Gurau}

\Address{CPHT, Ecole Polytechnique, 91128 Palaiseau cedex, France}
\Email{\href{mailto:rgurau@cpht.polytechnique.fr}{rgurau@cpht.polytechnique.fr}}
\URLaddress{\url{https://www.cpht.polytechnique.fr/?q=fr/node/85}}

\ArticleDates{Received September 21, 2016; Published online September 23, 2016}

\Abstract{This article is preface to the SIGMA special issue ``Tensor Models, Formalism and Applications'',
\url{http://www.emis.de/journals/SIGMA/Tensor_Models.html}. The issue is a collection of eight excellent, up to date reviews on random tensor models. The reviews combine pedagogical introductions meant for a general audience with presentations of the most recent developments in the f\/ield. This preface aims to give a condensed panoramic overview of random tensors as the natural generalization
of random matrices to higher dimensions.}


\Classification{83C45}

\renewcommand{\thefootnote}{\arabic{footnote}}
\setcounter{footnote}{0}

\pdfbookmark[1]{Why random tensors?}{section}
\section*{Why random tensors?}

General relativity, or classical gravity, is a theory of the ambient space-time geometry. The electroweak and strong interactions, that is the other three fundamental forces in nature, are described by perturbatively renormalizable quantum f\/ield theory \cite{Glashow1, Gross:1973ju,Gross:1974cs,Gross:1973id,Politzer:1973fx,Salam1,'tHooft:1972fi,Weinberg1} living on this geometric background. The main lesson of general relativity is that the ambient geometry is dynamical, and the main lesson of quantum f\/ield theory is that all the dynamical f\/ields must be quantized. Thus, classical gravity predicts its own demise: a more fundamental theory, ``quantum gravity'', must come into play at some high energy scale.

While classical general relativity is a f\/ield theory, it cannot be quantized the same way the standard model is: general relativity is \emph{not} perturbatively renormalizable \cite{sagnotti, tHooftgrav}. The lack of perturbative renormalizability of general relativity is a clear indicator that ``quantum gravity'' is quite dif\/ferent from the classical theory of gravity. This is why minimalistic approaches should be taken with a grain of salt: gravity is weakly coupled in the infrared, hence strongly coupled in the ultraviolet. This suggest that the fundamental degrees of freedom of quantum gravity are quite dif\/ferent from the geometric degrees of freedom perceived by low energy observers. It is far more likely that the geometric infrared degrees of freedom are just bound states of the genuine quantum gravity ultraviolet degrees of freedom.

Over the years several candidate quantum gravity theories have been developed, most notably string theory. While much remains to be learned about the elusive fundamental theory of quantum gravity, one thing is certain: whatever this theory may be, it must make sense of an expression like:
\begin{gather}
\label{eq:qg}
 \sum_{\text{topologies}} \int [{\cal D} g] [dX] \, e^{-S_{\rm EH}- S_{\rm SM}-S_{\text{other}}} .
\end{gather}
This is how quantum gravity should look like \cite{DiFrancesco:1993nw}: a path integral over matter f\/ields~$X$ and over geometric degrees of freedom (for instance metrics $g$ and topologies) of the exponential of the Einstein--Hilbert action plus the standard model action plus possibly other terms, describing physics yet to be discovered. The sum over topologies in equation~\eqref{eq:qg} can be reduced to a single topology by f\/ine tuning $S_{\text{other}}$, but \emph{a priori} there is no reason to do this: one lesson of quantum f\/ield theory is that whatever can happen will happen, and this includes topology change.

As already hinted, the precise meaning of equation~\eqref{eq:qg} is unclear: the sum over topologies and metrics is ill def\/ined and the naive attempts to use this equation as the starting point for a~theory of quantum gravity fail. A strategy to make sense of this equation is to replace the sum over (continuum) geometric degrees of freedom with a sum over triangulations:
\begin{gather*}
 \sum_{\text{topologies} } \int [{\cal D} g] \quad \rightarrow \quad \sum_{\text{triangulations}} ,
\end{gather*}
and to replace the continuum actions by their discretized versions. However, when passing to a~discrete setting, several questions arise, most notably:
\begin{itemize}\itemsep=0pt
\item what is the weight (probability distribution) one should use to sum over triangulations?
\item how does one go back from discrete to continuum geometries?
\end{itemize}

The answers to these questions are intertwined. In order to go back from discrete to continuum geometries one needs some kind of phase transition. The precise features of this phase transition strongly depend on the probability distribution chosen. In contrast to two or three dimensions, in dimension four information about the smooth structure is lost when going to a~discretization. Some of this information could perhaps index the available phases of continuous geometry.

Geometry and topology become increasingly complicated when increasing the topological dimension and it seems reasonable to f\/irst address these questions in dimension two, and only afterwards pass to the more realistic dimensions three or four.

In two dimensions the answers to these questions are provided by the theory of random matrices \cite{Gui,DiFrancesco:1993nw,Mehta, wigner,wishart}. Random matrices are probability distributions for $N\times N$ random variables $\mathbb{M}_{ab}$, which are invariant under the conjugation of $\mathbb{M}$ by the unitary group. The moments and partition function of a random matrix model can be evaluated as sums over ribbon Feynman graphs, with weights f\/ixed by the Feynman rules. These ribbon graphs are dual to topological surfaces. As the probability distribution of the
surfaces (weights of the graphs) is f\/ixed by the Feynman rules, random matrices yield a \emph{canonical} theory of random two-dimensional topological surfaces and provide an answer to the f\/irst question. One still has some freedom in assigning metrics to the triangulated topological surfaces. The simplest choice is to consider the triangulations as equilateral.

As always in quantum f\/ield theory, the perturbative Feynman series diverges. However, the situation is much more subtle in matrix models than in usual quantum f\/ield theory. A matrix model is endowed with a natural small parameter, $1/N$ (where $N$ is the size of the matrix), which does not exist in usual quantum f\/ield theory, and one can reorganize the perturbative expansion of a matrix model as a series in~$1/N$. In his seminal work~\cite{'tHooft:1973jz} 't~Hooft showed that the~$1/N$ series is indexed by the genus. This is the fundamental feature of matrix models: the~$1/N$ expansion reorganizes the perturbative series into nontrivial but manageable packages of graphs of f\/ixed genus.

At leading order in $1/N$ the planar graphs \cite{Brezin:1977sv,David:1984tx} dominate. While planar graphs can be arbitrarily large (i.e., they can have an arbitrary number of edges), they can be explicitly enumerated \cite{Gilles3,Gilles2,Gilles1} and form an exponentially bounded family. This holds order by order in $1/N$: the family of graphs of any f\/ixed genus is exponentially bounded. Tuning the coupling constants of a matrix model to some critical values, inf\/inite graphs with f\/ixed genus will dominate at any order in~$1/N$. At this critical point the model undergoes a phase transition to a continuum theory of random, inf\/initely ref\/ined, surfaces~\cite{David:1988hj,Kazakov:1985ds}. This answers the second question above. Assuming an equilateral metric assignment, the inf\/initely ref\/ined random geometry emerging at criticality, the Brownian map~\cite{BrownMap1,BrownMap2,BrownMap3}, has average Hausdorf\/f dimension 4 and is widely believed to have average spectral dimension~2.

The behavior of conformal matter coupled to Liouville gravity in two dimensions is very well captured by matrix models \cite{Ambjorn:1990ji,Boulatov:1986sb,Brezin:1989db, David:1988hj,DiFrancesco:1993nw,Dijkgraaf:1990rs, DK,Dup, Fukuma:1990jw,Kazakov:1986hu,Kazakovmulticrit, Knizhnik:1988ak,Makeenko:1991ry}. The double scaling limit of matrix models \cite{Brezin:1990rb,double1,double2} corresponds to a continuum gravity theory with f\/inite renormalized Newton's constant. Matrix models extend naturally to matrix f\/ield theories, which are intimately related to noncommutative quantum f\/ield theories \cite{GW,GW1}, in particular to the Grosse Wulkenhaar model, which has been shown asymptotically safe at all orders \cite{GW2} and solved in the planar sector \cite{GW3, GW4,GW5,GW6}.

Matrix models can be studied using the eigenvalue decomposition or the Schwinger--Dyson equations \cite{Guilarge,Eynard1,Eynard3,Eynard,Guilarge1,Eynard2}, avoiding the divergent perturbative expansion. However, the link with random surfaces must be revisited in this case. In order to obtain a theory of random surfaces one f\/irst performs the perturbative expansion and subsequently reorganizes it in powers of $1/N$. As the perturbative expansion is not summable, it is a non trivial task~\cite{thomasmatrix} to establish the connection between random surfaces and matrix models rigorously. Even more tantalizing, the critical point where the continuum limit is reached is
at \emph{negative} values of the coupling constants, in a range of parameters where the models are unstable. This is a feature, not an accident: the critical point must correspond to a regime where all the surfaces add up, and not to a point where the sum over surfaces is alternated. The tuning to criticality is meaningful after restricting to a f\/ixed order in~$1/N$, but what is the meaning of this tuning to criticality beyond perturbation theory?

Although many non perturbative questions about random matrices are still open, by and large, matrix models are a success story. One lesson can be drawn from their example: in two dimensions, in most cases, \emph{minimal choices suffice}. The relation between random matrices and quantum gravity
is already apparent if one studies the simplest matrix model and considers the simplest metric assignment (equilateral) for the triangulations.

The development of matrix models over the past decades is one of the most impressive achievements of modern theoretical and mathematical physics.
This success inspired their generalization in the 1990s to random tensor models \cite{ambj3dqg,mmgravity,sasa1,sasac,sasab} intended to describe random geometries in higher dimensions. However, for twenty years random tensors essentially failed to match the success of random matrices because, for a long time, a $1/N$ expansion for tensors could not be found.

Random tensors generate Feynman graphs that can be interpreted as topological spaces. However, the spaces generated in this way are quite nontrivial: one obtains not only all the manifolds, but also all the pseudo manifolds of a f\/ixed dimension\footnote{A further complication arose from the fact that the f\/irst models proposed did not generate genuine $D$-dimensional complexes, but only 2-complexes.}. Several models \cite{oldgft10,Baratin:2011aa,Boulatov:1992vp,oldgft6,Oriti:2011jm,Oriti:2014uga,Oriti:2014yla}, mainly under the guise of ``Group Field Theories'', which are tensor models decorated by extra data, have been proposed in the attempt to tackle this problem. Although dif\/ferent in some important respects, these models did not bring any insight into the problem of the $1/N$ expansion.

Starting in 2009 \cite{lost,PolyColor, color}, new results and techniques led to the discovery of the $1/N$ expansion for tensors \cite{uncoloring, expansion1,expansion3,expansion2}. These results form the backbone of the modern theory of random tensors. In this modern point of view, random tensors are probability distributions for~$N^D$ random variables $\mathbb{T}_{a_1 \dots a_D}$, which are invariant under the independent action of the unitary group on each tensor index:
\begin{gather*}
 \mathbb{T}_{a_1 \dots a_D} \to \mathbb{T}'_{b_1 \dots b_D} = \sum_{a_1,\dots a_D} U^{(1)}_{b_1 a_1} \cdots U^{(D)}_{b_D a_D} \mathbb{T}_{a_1 \dots a_D}.
\end{gather*}
The main building blocks of a random tensor model are the invariants one can build out of the tensor $ \mathbb{T} $ and its complex conjugate.
While for matrices there is essentially one such invariant at any degree, $\Tr[ ( {\mathbb M} {\mathbb M}^{\dagger})^p]$, there are numerous possible choices for tensors. Each of these invariants (generalized traces) can be represented by a bipartite $D$-regular edge colored graph, and there are as many independent invariants at degree $2p$ as there are non isomorphic such graphs with $2p$ vertices. This is the source of all the richness of random tensors.

The graphs of the new tensor models always encode genuine $D$-dimensional cellular complexes hence $D$-dimensional topological spaces. The $1/N$ series \cite{expansion4,expansioin5,expansion1,expansion3,expansioin6,expansion2} is indexed by a positive integer called the \emph{degree}, which plays in higher dimensions the same role the genus played for matrix models. Unlike the genus, however, the degree is \emph{not} a topological invariant, but mixes topological and triangulation dependent information. This is not a drawback, but a quality. Topology is quite complicated in dimensions three and four (or higher): although topologies are not fully classif\/ied in arbitrary dimension, the graphs at f\/ixed degree can be~\cite{GurSch}.

In the large $N$ limit the graphs of degree zero, called \emph{melonic} \cite{critical,universality,melbp}, dominate. The melonic graphs triangulate the $D$-dimensional sphere in any dimension \cite{expansion1,expansion3,expansion2} and are an exponentially bounded family. Like matrix models, tensor models undergo a phase transition to a theory of continuous random $D$-dimensional topological spaces \cite{critical,uncoloring} when tuning to criticality and posses a double scaling limit \cite{Bonzom:2014oua,DGR, GurSch}. In stark contrast to matrix models, for $3\le D\le 5$, in the double scaling regime only an exponentially bounded family of triangulations of the sphere contribute. For $D\ge 6$ (as for $D=2$) the family selected in the double scaling limit is neither exponentially bounded nor restricted to the spherical topology. The $1/N$ expansion of tensor models exhibits very strong universality properties \cite{sdequations,sdequations1, universality} and has been put on a f\/irm mathematical footing~\cite{expansioin6} in the sense of constructive f\/ield theory~\cite{GlimmJaffe} by means of the loop vertex expansion \cite{LVE4,LVE3, LVE1,LVE2}. The ensuing theory of random geometries in higher dimensions has been extensively studied \cite{Geloun:2013zka,Geloun:2014kpa,sefu2,Geloun:2013kta,EDT,Bonzom:2012np,sdequations3,Bonzom:2012qx,IsingD,Carrozza:2011jn,Carrozza:2012kt,
Fusy:2014rba,doubletens,sdequations,sdequations1,review,Gurau:2015tua,sdequations2,Ryan:2011qm,Tanasa:2011ur}.

Random tensor models generalize random matrix models and provide a framework for the study of random geometries in any dimension. However, as gravity is quite peculiar in two dimensions (and it is peculiar in a dif\/ferent way in three dimensions, and it is peculiar in yet another dif\/ferent way in four), one should keep an open mind for alternatives.

In their simplest form, tensor models can be seen, like matrix models, as generators of Euclidean dynamical triangulations \cite{EDTAmbjorn,EDTDavid, RTM}. There exists a tensor model \cite{critical} whose free energy $W_{\rm TM}(\lambda,N)$ equals an Einstein--Hilbert gravity partition function discretized on an equilateral triangulation with edge length~$a$:
\begin{gather*}
 \sum_{\text{topologies} } \int [{\cal D} g] \, e^{ \frac{1}{16 \pi G} \int d^Dx \sqrt{g} (R -2\Lambda) }
 \rightarrow \sum_{\genfrac{}{}{0pt}{}{ \text{Triangulations}}{\text{edge length } a}} e^{ -S_{\rm EH}^{\rm discr.} (G,\Lambda;a) } = W_{\rm TM }(\lambda,N) ,
\end{gather*}
where the coupling constant $\lambda$ and the parameter $N$ of the tensor model are related to the edge length $a$, the dimensionfull cosmological constant $\Lambda$ and Newton's constant $G$ respectively the dimensionless cosmological constant $\tilde \Lambda$ and Newton's constant $\tilde G$
by the relations:
\begin{gather*}
 \tilde G = \frac{G }{a^{D-2}} = c_1 \frac{1}{\ln N} , \qquad \tilde \Lambda = \Lambda a^2 = c_2 \frac{ G }{a^{D-2}} \ln \left(\frac{1}{\lambda} \right) + c_3 = \tan \alpha \tilde G + c_3 ,
\end{gather*}
with $c_1,c_2 , c_3 > 0$ some constants of order $1$ (and $c_3=0$ for $D=2$) and $\tan\alpha = c_2 \ln \big(\frac{1}{\lambda} \big) $.

\begin{figure}[t]\centering
\includegraphics{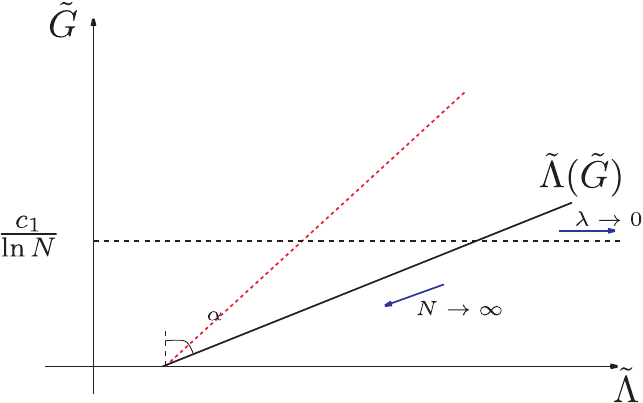}
\caption{The $(\tilde G,\tilde \Lambda)$ plane.}\label{fig:Glambda}
\end{figure}

The $(\tilde G,\tilde \Lambda)$ plane is represented in Fig.~\ref{fig:Glambda}. The angle $\alpha$ is the \emph{slope} of the (solid) line $\tilde \Lambda(\tilde G) = \tan \alpha \tilde G + c_3$ with respect to the vertical axis. The coupling constant $\lambda$ controls $\alpha$, while $N$ controls the height of the dashed line in Fig.~\ref{fig:Glambda}. The physical constants $\tilde G$ and $\tilde \Lambda$ are the coordinates of the intersection point of the two lines. For $\lambda \in (0,\infty)$, $\alpha \in \big(\frac{\pi}{2}, -\frac{\pi}{2}\big)$ and, as $N>1$, the two parameters $\lambda$ and $N$ allow one to explore the entire upper half plane of the physical constants.

As long as $a$ is kept f\/ixed, the dimensionfull constants are just a f\/inite rescaling of the dimensionless ones. The correspondence between the various regimes of tensor models and the physical coupling constants can be read out from this f\/igure, observing that $\lambda\to 0$ at f\/ixed~$N$ moves the intersection point to the right $\big(\alpha\to \frac{\pi}{2}\big)$ at constant height, while $N \to \infty$ at f\/ixed $\lambda$ descends the intersection point along the line $ \tilde \Lambda(\tilde G)$ at f\/ixed slope~$\alpha$.
\begin{description}\itemsep=0pt
 \item[\it The perturbative expansion.]
We have $a$ and $N$ f\/ixed, $\lambda\to 0$. The perturbative expansion is an expansion at \emph{large cosmological constant}
 and \emph{fixed Newton's constant}:
 \begin{gather*}
 \tilde G \ \text{constant (small)}, \qquad \alpha \to \frac{\pi}{2} , \qquad \tilde \Lambda \to \infty .
 \end{gather*}
 This is an expansion around the zero volume state corresponding to $ \tilde \Lambda = \infty$.

 \item[\it The $1/N$ expansion.]
 We have $a$ and $\lambda$ f\/ixed, $N \to \infty$. For $D\ge 3$, the $1/N$ expansion is an expansion at \emph{finite cosmological constant},
 \emph{small Newton's constant} and constant slope:
 \begin{gather*}
 \tilde G \to 0 , \qquad \alpha \ \text{constant} , \qquad \tilde \Lambda \to c_3 .
 \end{gather*}
 This is an expansion around geometries of maximal positive curvature at f\/ixed volume. The situation is dif\/ferent in $D=2$. In that case the $1/N$ expansion is an expansion at \emph{small cosmological constant} and \emph{small Newton's constant} with constant slope.

 \item[\it The large $N$ limit.]
We have $a$, $\lambda$ f\/ixed, $N = \infty$. In this limit one has a \emph{finite cosmological constant} and \emph{zero Newton's constant}, but this regime is approached along a line with f\/ixed slope:
 \begin{gather*}
 \tilde G = 0 , \qquad \alpha \ \text{ constant}, \qquad \tilde \Lambda = c_3 .
 \end{gather*}
 The large $N$ limit projects onto geometries with maximal positive curvature at f\/ixed volume. Among these geometries, one has a second expansion governed by the slope $\alpha$ around the zero volume state. That is, once projected onto geometries with maximal positive curvature, the slope $\alpha$ plays the role of an ef\/fective cosmological constant which is large and positive.

\item[\it The critical regime.]
We have $a$ f\/ixed, $\lambda\to \lambda_c$, $N = \infty$. One still has a \emph{finite cosmological constant} and \emph{zero Newton's constant}, but the slope approaches a critical value:
 \begin{gather*}
 \tilde G = 0 , \qquad \alpha \to \alpha_c , \qquad \tilde \Lambda = c_3 .
 \end{gather*}
 We represented the line with critical slope $\alpha_c$ in red in Fig.~\ref{fig:Glambda}. The critical value $\alpha_c$ corresponds to a divergent number of $D$-simplices hence, as $a$ is kept f\/ixed, these geometries have inf\/inite volume.

 \item[\it The continuum limit.]
 We have $a\to 0$, $\lambda\to \lambda_c$, $ a^D (\lambda_c - \lambda)^{-1}$ f\/ixed, $N = \infty$. The number of $D$-simplices diverges but the physical volume is kept f\/ixed by sending $a$ to zero while keeping $ a^D (\lambda_c - \lambda)^{-1}$ f\/ixed. The dimensionless and dimensionfull constants have a dif\/ferent behavior:
 \begin{gather*}
 \tilde G = 0 , \qquad \alpha \to \alpha_c , \qquad \tilde \Lambda = c_3 \qquad \Big{\vert}\qquad G = 0 , \qquad \Lambda = \infty .
 \end{gather*}

\item[\it The double scaling regime.]
 We have $a$ f\/ixed, $\lambda\to \lambda_c$, $N \to \infty$, $N^{D-2} (\lambda_c - \lambda) $ f\/ixed. This regime corresponds to a~well def\/ined trajectory in the plane of physical coupling constants:
 \begin{gather*}
 \tilde G \to 0 , \qquad \alpha \to \alpha_c , \qquad \tilde \Lambda \to c_3 .
 \end{gather*}

 \item[\it The continuum double scaling limit.]
 We have $a\to 0$, $\lambda\to \lambda_c$, $N \to \infty$, with $ N^{D-2} (\lambda_c - \lambda) $ f\/ixed and $ a^D (\lambda_c - \lambda)^{-1}$ f\/ixed. In this regime we have:
 \begin{gather*}
 \tilde G \to 0 , \qquad \alpha \to \alpha_c , \qquad \tilde \Lambda \to c_3 \qquad \Big{|} \qquad G \to 0 , \qquad \Lambda \to \infty .
 \end{gather*}
\end{description}

The analytically accessible regimes of this tensor model explore regions with \emph{small positive $($or zero$)$} Newton constant and \emph{large $($or very large$)$ positive} cosmological constant. Care should be taken when interpreting these values. This theory is supposed to describe the ultraviolet behavior of gravity, therefore the cosmological and Newton's constant we are discussing here are the ultraviolet ones. In order to obtain the infrared coupling constants one needs to follow a~renormalization group f\/low, and both $\Lambda$ and $G$ will vary substantially as they are dimensionfull in $D\ge 3$.

The geometries of maximal positive curvature at f\/ixed volume are rather simple and fall into the universality class of the continuous random tree (branched polymers). On the one hand this is good news, as one can treat analytically statistical systems in random geometry \cite{EDT,Bonzom:2012np,MeandersBonzom,Bonzom:2012qx,IsingD}, hence the coupling of gravity with matter f\/ields. On the other, one can only do so much with branched polymers and it is important to f\/ind regimes in which the emergent geometries are richer.

The double scaling limit is one attempt to go beyond branched polymers. In this limit larger families of graphs are included and, more importantly, below $D=6$ it seems likely that the double scaling regime can be followed by a triple scaling regime, revealing an even larger family of graphs. The strategy one employs in this context is to keep the metric interpretation of a~graph as dual to an equilateral triangulation, and to extend (in a~controlled manner) the family of graphs included in the statistical ensemble.

A second option is to encode metric degrees of freedom in additional data associated to Feynman graphs. This strategy is sometimes called group f\/ield theory (GFT) \cite{Baratin:2013rja,oldgft10,Baratin:2011aa,Boulatov:1992vp, Kegeles:2016wfg,oldgft6,Oriti:2011jm,Oriti:2014uga,Oriti:2014yla} and has proven quite successful. In GFT one considers tensors over some Lie group that are furthermore invariant under the diagonal action of the group. With minimal adaptations the $1/N$ expansion holds for GFTs~\cite{expansion1,expansion3}. Due to the diagonal gauge invariance, the amplitude of a GFT graph is the discretized BF action on the dual triangulation. At least in three dimensions, this is exactly the gravity partition function on the dual triangulation (in higher dimensions the BF action must be supplemented by constraints). The tantalizing fact is that, choosing for instance the group ${\rm SU}(2)$ in three dimensions, the melonic triangulations (which still dominate) are endowed with a \emph{flat connection}. This, most def\/initely, is \emph{not} a~branched polymer geometry. However, the reader should be aware of the following fact. While a~metric on a (pseudo) manifold is uniquely encoded in the holonomies around all the closed loops in that (pseudo) manifold, when discretizing one loses a \emph{local conformal factor}. While the fact that melonic geometries have a f\/lat connection is very encouraging, before f\/ixing this conformal factor issue one cannot control the emerging random geometry in GFTs. Be that as it may, GFT has successfully been applied to quantum cosmology \cite{Gielen:2015kua,Gielen:2014usa,Gielen:2016uft,Gielen:2014uga, Gielen:2013kla,Gielen:2013naa,Oriti:2015qva,Oriti:2015rwa,Oriti:2016ueo,Oriti:2016qtz,Sindoni:2014wya},
in what is the most relevant phenomenological application of tensor models to date.

A third option is to extend the framework of tensor models to tensor f\/ield theories. As matrix models extend naturally to matrix f\/ield theories, tensor models extend naturally to tensor f\/ield theories (TFT) \cite{Rivasseau:2011hm,Rivasseau:2012yp,Rivasseau:2013uca} by breaking the unitary invariance of the Gaussian part of the measure. The covariance of such models possesses a nontrivial spectrum, which in turn is naturally divided into scales. The integration of the high scales leads to a genuine renormalization group f\/low \cite{BenGeloun:2012yk, BenGeloun:2011rc,BenGeloun:2012pu} and the TFTs explore the \emph{tensor theory space}~\cite{Rivasseau:2014ima} spanned by all the unitary invariant polynomial interactions. One can combine both tensors with a~diagonal gauge invariance and a nontrivial covariance in what could be called tensor group f\/ield theory. Such models are also perturbatively renormalizable (which is highly nontrivial in this case, and relies crucially on the ultraviolet dominance of the melonic graphs), and exhibit a very rich phase portrait. While much remains to be done, one could hope that this class of models can f\/ix the local conformal factor problem of the usual GFTs.

The TFTs are examples of \emph{renormalizable, nonlocal} f\/ield theories. A regime parallel to the large $N$ limit of tensor models
is reached in TFTs via a genuine renormalization group f\/low \cite{BenGeloun:2012yk,jbgelon,Geloun:2014kpa,Geloun:2016xep,Geloun:2012bz,Geloun:2016qyb,Geloun:2013kta,BenGeloun:2011rc,BenGeloun:2012pu,Geloun:2014ema,Benedetti:2014qsa,
Benedetti:2015yaa,Carrozza:2014rba,Carrozza:2014rya,tt2,Carrozza:2012uv,
Lahoche:2015tqa,Lahoche:2016xiq, Samary:2013xla,Samary:2014tja, Samary:2012bw}, and not by sending the parameter $N$ to inf\/inity. The large $N$ behavior of tensor models corresponds to the ultraviolet limit of TFTs. Typically the couplings grow in the infrared which suggests that TFTs develop bound states at low energy. This low energy behavior remains largely to be explored either by analytic \cite{Carrozza:2014rba,Carrozza:2014rya} or by numeric \cite{Benedetti:2014qsa} methods. We emphasize that the natural metric assignment for the triangulations dual to graphs in TFTs is not yet understood. However, the fact that TFTs develop bound states in the infrared is an encouraging sign and establish them as potential ultraviolet completions of gravity.

Random tensors are the straightforward generalization of random matrices in higher dimensions. It should however be stressed that the $D=2$ case of matrices is very special. Indeed, random tensors behave by and large quite dif\/ferently from random matrices. This is due to the fact that the melonic family, which dominates in tensor models, is very dif\/ferent from (and in fact much more restricted than) the planar family dominating matrix models.

\pdfbookmark[1]{References}{ref}
\LastPageEnding


\begin{thebibliography}{99}
\footnotesize\itemsep=-0.5pt

\bibitem{ambj3dqg}
Ambj{\o}rn J., Durhuus B., J{\'o}nsson T., Three-dimensional simplicial quantum
 gravity and generalized matrix models, \href{http://dx.doi.org/10.1142/S0217732391001184}{\textit{Modern Phys. Lett.~A}}
 \textbf{6} (1991), 1133--1146.

\bibitem{EDTAmbjorn}
Ambj{\o}rn J., Durhuus B., Jonsson T., Quantum geometry. A~statistical f\/ield
 theory approach, \href{http://dx.doi.org/10.1017/CBO9780511524417}{\textit{Cambridge Monographs on Mathematical Physics}}, Cambridge
 University Press, Cambridge, 1997.

\bibitem{Ambjorn:1990ji}
Ambj{\o}rn J., Jurkiewicz J., Makeenko Yu.M., Multiloop correlators for
 two-dimensional quantum gravity, \href{http://dx.doi.org/10.1016/0370-2693(90)90790-D}{\textit{Phys. Lett.~B}} \textbf{251} (1990),
 517--524.

\bibitem{Gui}
Anderson G.W., Guionnet A., Zeitouni O., An introduction to random matrices,
 \textit{Cambridge Studies in Advanced Mathematics}, Vol.~118, Cambridge
 University Press, Cambridge, 2010.

\bibitem{Baratin:2013rja}
Baratin A., Carrozza S., Oriti D., Ryan J., Smerlak M., Melonic phase
 transition in group f\/ield theory, \href{http://dx.doi.org/10.1007/s11005-014-0699-9}{\textit{Lett. Math. Phys.}} \textbf{104}
 (2014), 1003--1017, \href{http://arxiv.org/abs/1307.5026}{arXiv:1307.5026}.

\bibitem{oldgft10}
Baratin A., Oriti D., Group f\/ield theory with noncommutative metric variables,
 \href{http://dx.doi.org/10.1103/PhysRevLett.105.221302}{\textit{Phys. Rev. Lett.}} \textbf{105} (2010), 221302, 4~pages,
 \href{http://arxiv.org/abs/1002.4723}{arXiv:1002.4723}.

\bibitem{Baratin:2011aa}
Baratin A., Oriti D., Ten questions on group f\/ield theory (and their tentative
 answers, \href{http://dx.doi.org/10.1088/1742-6596/360/1/012002}{\textit{J.~Phys. Conf. Ser.}} \textbf{360} (2012), 012002, 10~pages,
 \href{http://arxiv.org/abs/1112.3270}{arXiv:1112.3270}.

\bibitem{Guilarge}
Ben~Arous G., Guionnet A., Large deviations for {W}igner's law and
 {V}oiculescu's non-commutative entropy, \href{http://dx.doi.org/10.1007/s004400050119}{\textit{Probab. Theory Related
 Fields}} \textbf{108} (1997), 517--542.

\bibitem{BenGeloun:2012yk}
Ben~Geloun J., Two- and four-loop {$\beta$}-functions of rank-4 renormalizable
 tensor f\/ield theories, \href{http://dx.doi.org/10.1088/0264-9381/29/23/235011}{\textit{Classical Quantum Gravity}} \textbf{29} (2012),
 235011, 40~pages, \href{http://arxiv.org/abs/1205.5513}{arXiv:1205.5513}.

\bibitem{jbgelon}
Ben~Geloun J., Asymptotic freedom of rank 4 tensor group f\/ield theory, in
 Symmetries and Groups in Contemporary Physics, \href{http://dx.doi.org/10.1142/9789814518550_0049}{\textit{Nankai Ser. Pure Appl.
 Math. Theoret. Phys.}}, Vol.~11, World Sci. Publ., Hackensack, NJ, 2013,
 367--372, \href{http://arxiv.org/abs/1210.5490}{arXiv:1210.5490}.

\bibitem{Geloun:2013zka}
Ben~Geloun J., On the f\/inite amplitudes for open graphs in {A}belian dynamical
 colored {B}oulatov--{O}oguri models, \href{http://dx.doi.org/10.1088/1751-8113/46/40/402002}{\textit{J.~Phys.~A: Math. Theor.}}
 \textbf{46} (2013), 402002, 12~pages, \href{http://arxiv.org/abs/1307.8299}{arXiv:1307.8299}.

\bibitem{Geloun:2014kpa}
Ben~Geloun J., Renormalizable models in rank {$d\geq 2$} tensorial group f\/ield
 theory, \href{http://dx.doi.org/10.1007/s00220-014-2142-6}{\textit{Comm. Math. Phys.}} \textbf{332} (2014), 117--188,
 \href{http://arxiv.org/abs/1306.1201}{arXiv:1306.1201}.

\bibitem{Geloun:2016xep}
Ben~Geloun J., Koslowski T.A., Nontrivial {UV} behavior of rank-4 tensor f\/ield
 models for quantum gravity, \href{http://arxiv.org/abs/1606.04044}{arXiv:1606.04044}.

\bibitem{Geloun:2012bz}
Ben~Geloun J., Livine E.R., Some classes of renormalizable tensor models,
 \href{http://dx.doi.org/10.1063/1.4818797}{\textit{J.~Math. Phys.}} \textbf{54} (2013), 082303, 25~pages,
 \href{http://arxiv.org/abs/1207.0416}{arXiv:1207.0416}.

\bibitem{sefu2}
Ben~Geloun J., Magnen J., Rivasseau V., Bosonic colored group f\/ield theory,
 \href{http://dx.doi.org/10.1140/epjc/s10052-010-1487-z}{\textit{Eur. Phys.~J.~C Part. Fields}} \textbf{70} (2010), 1119--1130,
 \href{http://arxiv.org/abs/0911.1719}{arXiv:0911.1719}.

\bibitem{Geloun:2016qyb}
Ben~Geloun J., Martini R., Oriti D., Functional renormalization group analysis
 of tensorial group f\/ield theories on~$\mathbb{R}^d$, \href{http://dx.doi.org/10.1103/PhysRevD.94.024017}{\textit{Phys. Rev.~D}}
 \textbf{94} (2016), 024017, 45~pages, \href{http://arxiv.org/abs/1601.08211}{arXiv:1601.08211}.

\bibitem{Geloun:2013kta}
Ben~Geloun J., Ramgoolam S., Counting tensor model observables and branched
 covers of the 2-sphere, \href{http://dx.doi.org/10.4171/AIHPD/4}{\textit{Ann. Inst. Henri Poincar\'e~D}} \textbf{1}
 (2014), 77--138, \href{http://arxiv.org/abs/1307.6490}{arXiv:1307.6490}.

\bibitem{BenGeloun:2011rc}
Ben~Geloun J., Rivasseau V., A renormalizable 4-dimensional tensor f\/ield
 theory, \href{http://dx.doi.org/10.1007/s00220-012-1549-1}{\textit{Comm. Math. Phys.}} \textbf{318} (2013), 69--109,
 \href{http://arxiv.org/abs/1111.4997}{arXiv:1111.4997}.

\bibitem{BenGeloun:2012pu}
Ben~Geloun J., Samary D.O., 3{D} tensor f\/ield theory: renormalization and
 one-loop {$\beta$}-functions, \href{http://dx.doi.org/10.1007/s00023-012-0225-5}{\textit{Ann. Henri Poincar\'e}} \textbf{14}
 (2013), 1599--1642, \href{http://arxiv.org/abs/1201.0176}{arXiv:1201.0176}.

\bibitem{Geloun:2014ema}
Ben~Geloun J., Toriumi R., Parametric representation of rank {$d$} tensorial
 group f\/ield theory: {A}belian models with kinetic term {$\sum_s\vert p_s\vert
 +\mu$}, \href{http://dx.doi.org/10.1063/1.4929771}{\textit{J.~Math. Phys.}} \textbf{56} (2015), 093503, 53~pages,
 \href{http://arxiv.org/abs/1409.0398}{arXiv:1409.0398}.

\bibitem{Benedetti:2014qsa}
Benedetti D., Ben~Geloun J., Oriti D., Functional renormalisation group
 approach for tensorial group f\/ield theo\-ry: a rank-3 model, \href{http://dx.doi.org/10.1007/JHEP03(2015)084}{\textit{J.~High
 Energy Phys.}} \textbf{2015} (2015), no.~3, 084, 40~pages, \href{http://arxiv.org/abs/1411.3180}{arXiv:1411.3180}.

\bibitem{EDT}
Benedetti D., Gurau R., Phase transition in dually weighted colored tensor
 models, \href{http://dx.doi.org/10.1016/j.nuclphysb.2011.10.015}{\textit{Nuclear Phys.~B}} \textbf{855} (2012), 420--437,
 \href{http://arxiv.org/abs/1108.5389}{arXiv:1108.5389}.

\bibitem{Benedetti:2015yaa}
Benedetti D., Lahoche V., Functional renormalization group approach for
 tensorial group f\/ield theo\-ry: a rank-6 model with closure constraint,
 \href{http://dx.doi.org/10.1088/0264-9381/33/9/095003}{\textit{Classical Quantum Gravity}} \textbf{33} (2016), 095003, 35~pages,
 \href{http://arxiv.org/abs/1508.0638}{arXiv:1508.0638}.

\bibitem{Bonzom:2012np}
Bonzom V., Multi-critical tensor models and hard dimers on spherical random
 lattices, \href{http://dx.doi.org/10.1016/j.physleta.2012.12.022}{\textit{Phys. Lett.~A}} \textbf{377} (2013), 501--506,
 \href{http://arxiv.org/abs/1201.1931}{arXiv:1201.1931}.

\bibitem{expansion4}
Bonzom V., New {$1/N$} expansions in random tensor models, \href{http://dx.doi.org/10.1007/JHEP06(2013)062}{\textit{J.~High
 Energy Phys.}} \textbf{2013} (2013), no.~6, 062, 25~pages, \href{http://arxiv.org/abs/1211.1657}{arXiv:1211.1657}.

\bibitem{sdequations3}
Bonzom V., Revisiting random tensor models at large {$N$} via the
 {S}chwinger--{D}yson equations, \href{http://dx.doi.org/10.1007/JHEP03(2013)160}{\textit{J.~High Energy Phys.}} \textbf{2013}
 (2013), no.~3, 160, 25~pages, \href{http://arxiv.org/abs/1208.6216}{arXiv:1208.6216}.

\bibitem{MeandersBonzom}
Bonzom V., Combes F., Tensor models from the viewpoint of matrix models: the
 cases of loop mo\-dels on random surfaces and of the {G}aussian distribution,
 \href{http://dx.doi.org/10.4171/AIHPD/14}{\textit{Ann. Inst. Henri Poincar\'e~D}} \textbf{2} (2015), 1--47,
 \href{http://arxiv.org/abs/1304.4152}{arXiv:1304.4152}.

\bibitem{Bonzom:2012qx}
Bonzom V., Erbin H., Coupling of hard dimers to dynamical lattices via random
 tensors, \href{http://dx.doi.org/10.1088/1742-5468/2012/09/P09009}{\textit{J.~Stat. Mech. Theory Exp.}} \textbf{2012} (2012), P09009,
 18~pages, \href{http://arxiv.org/abs/1204.3798}{arXiv:1204.3798}.

\bibitem{critical}
Bonzom V., Gurau R., Riello A., Rivasseau V., Critical behavior of colored
 tensor models in the large~{$N$} limit, \href{http://dx.doi.org/10.1016/j.nuclphysb.2011.07.022}{\textit{Nuclear Phys.~B}} \textbf{853}
 (2011), 174--195, \href{http://arxiv.org/abs/1105.3122}{arXiv:1105.3122}.

\bibitem{IsingD}
Bonzom V., Gurau R., Rivasseau V., The {I}sing model on random lattices in
 arbitrary dimensions, \href{http://dx.doi.org/10.1016/j.physletb.2012.03.054}{\textit{Phys. Lett.~B}} \textbf{711} (2012), 88--96,
 \href{http://arxiv.org/abs/1108.6269}{arXiv:1108.6269}.

\bibitem{uncoloring}
Bonzom V., Gurau R., Rivasseau V., Random tensor models in the large~$N$ limit:
 uncoloring the colored tensor models, \href{http://dx.doi.org/10.1103/PhysRevD.85.084037}{\textit{Phys. Rev.~D}} \textbf{85}
 (2012), 084037, 12~pages, \href{http://arxiv.org/abs/1202.3637}{arXiv:1202.3637}.

\bibitem{Bonzom:2014oua}
Bonzom V., Gurau R., Ryan J.P., Tanasa A., The double scaling limit of random
 tensor models, \href{http://dx.doi.org/10.1007/JHEP09(2014)051}{\textit{J.~High Energy Phys.}} \textbf{2014} (2014), no.~9,
 051, 49~pages, \href{http://arxiv.org/abs/1404.7517}{arXiv:1404.7517}.

\bibitem{Boulatov:1992vp}
Boulatov D.V., A model of three-dimensional lattice gravity, \href{http://dx.doi.org/10.1142/S0217732392001324}{\textit{Modern
 Phys. Lett.~A}} \textbf{7} (1992), 1629--1646, \href{http://arxiv.org/abs/hep-th/9202074}{hep-th/9202074}.

\bibitem{Boulatov:1986sb}
Boulatov D.V., Kazakov V.A., The {I}sing model on a random planar lattice: the
 structure of the phase transition and the exact critical exponents,
 \href{http://dx.doi.org/10.1016/0370-2693(87)90312-1}{\textit{Phys. Lett.~B}} \textbf{186} (1987), 379--384.

\bibitem{Brezin:1989db}
Br{\'e}zin {\'E}., Douglas M.R., Kazakov V., Shenker S.H., The {I}sing model
 coupled to {$2$}{D} gravity. {A}~nonperturbative analysis, \href{http://dx.doi.org/10.1016/0370-2693(90)90458-I}{\textit{Phys.
 Lett.~B}} \textbf{237} (1990), 43--46.

\bibitem{Brezin:1977sv}
Br{\'e}zin E., Itzykson C., Parisi G., Zuber J.B., Planar diagrams,
 \href{http://dx.doi.org/10.1007/BF01614153}{\textit{Comm. Math. Phys.}} \textbf{59} (1978), 35--51.

\bibitem{Brezin:1990rb}
Br{\'e}zin E., Kazakov V.A., Exactly solvable f\/ield theories of closed strings,
 \href{http://dx.doi.org/10.1016/0370-2693(90)90818-Q}{\textit{Phys. Lett.~B}} \textbf{236} (1990), 144--150.

\bibitem{Carrozza:2014rba}
Carrozza S., Discrete renormalization group for {${\rm SU}(2)$} tensorial group
 f\/ield theory, \href{http://dx.doi.org/10.4171/AIHPD/15}{\textit{Ann. Inst. Henri Poincar\'e~D}} \textbf{2} (2015),
 49--112, \href{http://arxiv.org/abs/1407.4615}{arXiv:1407.4615}.

\bibitem{Carrozza:2014rya}
Carrozza S., Group f\/ield theory in dimension {$4-\varepsilon$}, \href{http://dx.doi.org/10.1103/PhysRevD.91.065023}{\textit{Phys.
 Rev.~D}} \textbf{91} (2015), 065023, 10~pages, \href{http://arxiv.org/abs/1411.5385}{arXiv:1411.5385}.

\bibitem{Carrozza:2011jn}
Carrozza S., Oriti D., Bounding bubbles: the vertex representation of $3d$
 group f\/ield theory and the suppression of pseudomanifolds, \href{http://dx.doi.org/10.1103/PhysRevD.85.044004}{\textit{Phys.
 Rev.~D}} \textbf{85} (2012), 044004, 22~pages, \href{http://arxiv.org/abs/1104.5158}{arXiv:1104.5158}.

\bibitem{Carrozza:2012kt}
Carrozza S., Oriti D., Bubbles and jackets: new scaling bounds in topological
 group f\/ield theories, \href{http://dx.doi.org/10.1007/JHEP06(2012)092}{\textit{J.~High Energy Phys.}} \textbf{2012} (2012),
 no.~6, 092, 42~pages, \href{http://arxiv.org/abs/1203.5082}{arXiv:1203.5082}.

\bibitem{tt2}
Carrozza S., Oriti D., Rivasseau V., Renormalization of a {${\rm SU}(2)$}
 tensorial group f\/ield theory in three dimensions, \href{http://dx.doi.org/10.1007/s00220-014-1928-x}{\textit{Comm. Math. Phys.}}
 \textbf{330} (2014), 581--637, \href{http://arxiv.org/abs/1303.6772}{arXiv:1303.6772}.

\bibitem{Carrozza:2012uv}
Carrozza S., Oriti D., Rivasseau V., Renormalization of tensorial group f\/ield
 theories: {A}belian {${\rm U}(1)$} models in four dimensions, \href{http://dx.doi.org/10.1007/s00220-014-1954-8}{\textit{Comm.
 Math. Phys.}} \textbf{327} (2014), 603--641, \href{http://arxiv.org/abs/1207.6734}{arXiv:1207.6734}.

\bibitem{Gilles3}
Chapuy G., Marcus M., Schaef\/fer G., A bijection for rooted maps on orientable
 surfaces, \href{http://dx.doi.org/10.1137/080720097}{\textit{SIAM~J. Discrete Math.}} \textbf{23} (2009), 1587--1611,
 \href{http://arxiv.org/abs/0712.3649}{arXiv:0712.3649}.

\bibitem{Gilles2}
Cori R., Schaef\/fer G., Description trees and {T}utte formulas, \href{http://dx.doi.org/10.1016/S0304-3975(01)00221-3}{\textit{Theoret.
 Comput. Sci.}} \textbf{292} (2003), 165--183.

\bibitem{DGR}
Dartois S., Gurau R., Rivasseau V., Double scaling in tensor models with a
 quartic interaction, \href{http://dx.doi.org/10.1007/JHEP09(2013)088}{\textit{J.~High Energy Phys.}} \textbf{2013} (2013),
 no.~9, 088, 33~pages, \href{http://arxiv.org/abs/1307.5281}{arXiv:1307.5281}.

\bibitem{expansioin5}
Dartois S., Rivasseau V., Tanasa A., The {$1/N$} expansion of multi-orientable
 random tensor models, \href{http://dx.doi.org/10.1007/s00023-013-0262-8}{\textit{Ann. Henri Poincar\'e}} \textbf{15} (2014),
 965--984, \href{http://arxiv.org/abs/1301.1535}{arXiv:1301.1535}.

\bibitem{David:1984tx}
David F., Planar diagrams, two-dimensional lattice gravity and surface models,
 \href{http://dx.doi.org/10.1016/0550-3213(85)90335-9}{\textit{Nuclear Phys.~B}} \textbf{257} (1985), 45--58.

\bibitem{David:1988hj}
David F., Conformal f\/ield theories coupled to {$2$}-{D} gravity in the
 conformal gauge, \href{http://dx.doi.org/10.1142/S0217732388001975}{\textit{Modern Phys. Lett.~A}} \textbf{3} (1988), 1651--1656.

\bibitem{EDTDavid}
David F., Simplicial quantum gravity and random lattices, in Gravitation and
 Quantizations ({L}es {H}ouches, 1992), Editors J.~Zinn-Justin, B.~Julia,
 North-Holland, Amsterdam, 1995, 679--749, \href{http://arxiv.org/abs/hep-th/9303127}{hep-th/9303127}.

\bibitem{DiFrancesco:1993nw}
Di~Francesco P., Ginsparg P., Zinn-Justin J., {$2$}{D} gravity and random
 matrices, \href{http://dx.doi.org/10.1016/0370-1573(94)00084-G}{\textit{Phys. Rep.}} \textbf{254} (1995), 1--133,
 \href{http://arxiv.org/abs/hep-th/9306153}{hep-th/9306153}.

\bibitem{Dijkgraaf:1990rs}
Dijkgraaf R., Verlinde H., Verlinde E., Loop equations and {V}irasoro
 constraints in nonperturbative two-dimensional quantum gravity,
 \href{http://dx.doi.org/10.1016/0550-3213(91)90199-8}{\textit{Nuclear Phys.~B}} \textbf{348} (1991), 435--456.

\bibitem{GW2}
Disertori M., Gurau R., Magnen J., Rivasseau V., Vanishing of beta function of
 non-commutative {$\Phi_4^4$} theory to all orders, \href{http://dx.doi.org/10.1016/j.physletb.2007.04.007}{\textit{Phys. Lett.~B}}
 \textbf{649} (2007), 95--102, \href{http://arxiv.org/abs/hep-th/0612251}{hep-th/0612251}.

\bibitem{DK}
Distler J., Kawai H., Conformal f\/ield theory and {$2$}{D} quantum gravity,
 \href{http://dx.doi.org/10.1016/0550-3213(89)90354-4}{\textit{Nuclear Phys.~B}} \textbf{321} (1989), 509--527.

\bibitem{double1}
Douglas M.R., Shenker S.H., Strings in less than one dimension, \href{http://dx.doi.org/10.1016/0550-3213(90)90522-F}{\textit{Nuclear
 Phys.~B}} \textbf{335} (1990), 635--654.

\bibitem{Dup}
Duplantier B., Conformal random geometry, in \href{http://dx.doi.org/10.1016/S0924-8099(06)80040-5}{Mathematical Statistical Physics},
 Elsevier~B.V., Amsterdam, 2006, 101--217, \href{http://arxiv.org/abs/math-ph/0608053}{math-ph/0608053}.

\bibitem{Eynard1}
Eynard B., Topological expansion for the 1-{H}ermitian matrix model correlation
 functions, \href{http://dx.doi.org/10.1088/1126-6708/2004/11/031}{\textit{J.~High Energy Phys.}} \textbf{2004} (2004), no.~11, 031,
 35~pages, \href{http://arxiv.org/abs/hep-th/0407261}{hep-th/0407261}.

\bibitem{Eynard3}
Eynard B., Another algebraic variational principle for the spectral curve of
 matrix models, \href{http://arxiv.org/abs/1407.8324}{arXiv:1407.8324}.

\bibitem{Eynard}
Eynard B., Orantin N., Invariants of algebraic curves and topological
 expansion, \href{http://dx.doi.org/10.4310/CNTP.2007.v1.n2.a4}{\textit{Commun. Number Theory Phys.}} \textbf{1} (2007), 347--452,
 \href{http://arxiv.org/abs/math-ph/0702045}{math-ph/0702045}.

\bibitem{Fukuma:1990jw}
Fukuma M., Kawai H., Nakayama R., Continuum {S}chwinger--{D}yson equations and
 universal structures in two-dimensional quantum gravity, \href{http://dx.doi.org/10.1142/S0217751X91000733}{\textit{Internat.~J.
 Modern Phys.~A}} \textbf{6} (1991), 1385--1406.

\bibitem{Fusy:2014rba}
Fusy E., Tanasa A., Asymptotic expansion of the multi-orientable random tensor
 model, \textit{Electron.~J. Combin.} \textbf{22} (2015), 1.52, 30~pages,
 \href{http://arxiv.org/abs/1408.5725}{arXiv:1408.5725}.

\bibitem{Gielen:2015kua}
Gielen S., Identifying cosmological perturbations in group f\/ield theory
 condensates, \href{http://dx.doi.org/10.1007/JHEP08(2015)010}{\textit{J.~High Energy Phys.}} \textbf{2015} (2015), no.~8, 010,
 23~pages, \href{http://arxiv.org/abs/1505.0747}{arXiv:1505.0747}.

\bibitem{Gielen:2014usa}
Gielen S., Perturbing a quantum gravity condensate, \href{http://dx.doi.org/10.1103/PhysRevD.91.043526}{\textit{Phys. Rev.~D}}
 \textbf{91} (2015), 043526, 11~pages, \href{http://arxiv.org/abs/1411.1077}{arXiv:1411.1077}.

\bibitem{Gielen:2016uft}
Gielen S., Emergence of a low spin phase in group f\/ield theory condensates,
 \href{http://arxiv.org/abs/1604.06023}{arXiv:1604.06023}.

\bibitem{Gielen:2014uga}
Gielen S., Oriti D., Quantum cosmology from quantum gravity condensates:
 cosmological variables and lattice-ref\/ined dynamics, \href{http://dx.doi.org/10.1088/1367-2630/16/12/123004}{\textit{New~J. Phys.}}
 \textbf{16} (2014), 123004, 11~pages, \href{http://arxiv.org/abs/1407.8167}{arXiv:1407.8167}.

\bibitem{Gielen:2013kla}
Gielen S., Oriti D., Sindoni L., Cosmology from group f\/ield theory formalism
 for quantum gravity, \href{http://dx.doi.org/10.1103/PhysRevLett.111.031301}{\textit{Phys. Rev. Lett.}} \textbf{111} (2013), 031301,
 4~pages, \href{http://arxiv.org/abs/1303.3576}{arXiv:1303.3576}.

\bibitem{Gielen:2013naa}
Gielen S., Oriti D., Sindoni L., Homogeneous cosmologies as group f\/ield theory
 condensates, \href{http://dx.doi.org/10.1007/JHEP06(2014)013}{\textit{J.~High Energy Phys.}} \textbf{2014} (2014), no.~6, 013,
 69~pages, \href{http://arxiv.org/abs/1311.1238}{arXiv:1311.1238}.

\bibitem{Glashow1}
Glashow S.L., Partial-symmetries of weak interactions, \href{http://dx.doi.org/10.1016/0029-5582(61)90469-2}{\textit{Nuclear Phys.}}
 \textbf{22} (1961), 579--588.

\bibitem{GlimmJaffe}
Glimm J., Jaf\/fe A., Quantum physics. A functional integral point of view, 2nd~ed., \href{http://dx.doi.org/10.1007/978-1-4612-4728-9}{Springer-Verlag}, New York, 1987.

\bibitem{sagnotti}
Gorof\/f M.H., Sagnotti A., The ultraviolet behavior of {E}instein gravity,
 \href{http://dx.doi.org/10.1016/0550-3213(86)90193-8}{\textit{Nuclear Phys.~B}} \textbf{266} (1986), 709--736.

\bibitem{double2}
Gross D.J., Migdal A.A., Nonperturbative two-dimensional quantum gravity,
 \href{http://dx.doi.org/10.1103/PhysRevLett.64.127}{\textit{Phys. Rev. Lett.}} \textbf{64} (1990), 127--130.

\bibitem{Gross:1973ju}
Gross D.J., Wilczek F., Asymptotically free gauge theories.~I, \href{http://dx.doi.org/10.1103/PhysRevD.8.3633}{\textit{Phys.
 Rev.~D}} \textbf{8} (1973), 3633--3652.

\bibitem{Gross:1974cs}
Gross D.J., Wilczek F., Asymptotically free gauge theories.~II, \href{http://dx.doi.org/10.1103/PhysRevD.9.980}{\textit{Phys.
 Rev.~D}} \textbf{9} (1974), 980--993.

\bibitem{Gross:1973id}
Gross D.J., Wilczek F., Ultraviolet behavior of nonabelian gauge theories,
 \href{http://dx.doi.org/10.1103/PhysRevLett.30.1343}{\textit{Phys. Rev. Lett.}} \textbf{30} (1973), 1343--1346.

\bibitem{mmgravity}
Gross M., Tensor models and simplicial quantum gravity in {$>2$}-{D},
 \href{http://dx.doi.org/10.1016/S0920-5632(05)80015-5}{\textit{Nuclear Phys.~B Proc. Suppl.}} \textbf{25A} (1992), 144--149.

\bibitem{GW}
Grosse H., Wulkenhaar R., Renormalisation of {$\phi^4$}-theory on
 noncommutative {${\mathbb R}^4$} in the matrix base, \href{http://dx.doi.org/10.1007/s00220-004-1285-2}{\textit{Comm. Math.
 Phys.}} \textbf{256} (2005), 305--374, \href{http://arxiv.org/abs/hep-th/0401128}{hep-th/0401128}.

\bibitem{GW3}
Grosse H., Wulkenhaar R., Progress in solving a noncommutative quantum f\/ield
 theory in four dimensions, \href{http://arxiv.org/abs/0909.1389}{arXiv:0909.1389}.

\bibitem{GW4}
Grosse H., Wulkenhaar R., Solvable limits of a {$4D$} noncommutative {QFT},
 \href{http://arxiv.org/abs/1306.2816}{arXiv:1306.2816}.

\bibitem{GW5}
Grosse H., Wulkenhaar R., Construction of the {$\Phi^4_4$}-quantum f\/ield theory
 on noncommutative {M}oyal space, \textit{RIMS K\=oky\=uroku} \textbf{1904}
 (2013), 67--104, \href{http://arxiv.org/abs/1402.1041}{arXiv:1402.1041}.

\bibitem{GW6}
Grosse H., Wulkenhaar R., Solvable {4D} noncommutative {QFT}: phase transitions
 and quest for ref\/lection positivity, \href{http://arxiv.org/abs/1406.7755}{arXiv:1406.7755}.

\bibitem{Guilarge1}
Guionnet A., Zeitouni O., Concentration of the spectral measure for large
 matrices, \href{http://dx.doi.org/10.1214/ECP.v5-1026}{\textit{Electron. Comm. Probab.}} \textbf{5} (2000), 119--136.

\bibitem{lost}
Gurau R., Lost in translation: topological singularities in group f\/ield theory,
 \href{http://dx.doi.org/10.1088/0264-9381/27/23/235023}{\textit{Classical Quantum Gravity}} \textbf{27} (2010), 235023, 20~pages,
 \href{http://arxiv.org/abs/1006.0714}{arXiv:1006.0714}.

\bibitem{PolyColor}
Gurau R., Topological graph polynomials in colored group f\/ield theory,
 \href{http://dx.doi.org/10.1007/s00023-010-0035-6}{\textit{Ann. Henri Poincar\'e}} \textbf{11} (2010), 565--584,
 \href{http://arxiv.org/abs/0911.1945}{arXiv:0911.1945}.

\bibitem{expansion1}
Gurau R., The {$1/N$} expansion of colored tensor models, \href{http://dx.doi.org/10.1007/s00023-011-0101-8}{\textit{Ann. Henri
 Poincar\'e}} \textbf{12} (2011), 829--847, \href{http://arxiv.org/abs/1011.2726}{arXiv:1011.2726}.

\bibitem{color}
Gurau R., Colored group f\/ield theory, \href{http://dx.doi.org/10.1007/s00220-011-1226-9}{\textit{Comm. Math. Phys.}} \textbf{304}
 (2011), 69--93, \href{http://arxiv.org/abs/0907.2582}{arXiv:0907.2582}.

\bibitem{doubletens}
Gurau R., Double scaling limit in arbitrary dimensions: a~toy model,
 \href{http://dx.doi.org/10.1103/PhysRevD.84.124051}{\textit{Phys. Rev.~D}} \textbf{84} (2011), 124051, 11~pages,
 \href{http://arxiv.org/abs/1110.2460}{arXiv:1110.2460}.

\bibitem{sdequations}
Gurau R., A generalization of the {V}irasoro algebra to arbitrary dimensions,
 \href{http://dx.doi.org/10.1016/j.nuclphysb.2011.07.009}{\textit{Nuclear Phys.~B}} \textbf{852} (2011), 592--614, \href{http://arxiv.org/abs/1105.6072}{arXiv:1105.6072}.

\bibitem{expansion3}
Gurau R., The complete {$1/N$} expansion of colored tensor models in arbitrary
 dimension, \href{http://dx.doi.org/10.1007/s00023-011-0118-z}{\textit{Ann. Henri Poincar\'e}} \textbf{13} (2012), 399--423,
 \href{http://arxiv.org/abs/1102.5759}{arXiv:1102.5759}.

\bibitem{sdequations1}
Gurau R., The {S}chwinger--{D}yson equations and the algebra of constraints of
 random tensor models at all orders, \href{http://dx.doi.org/10.1016/j.nuclphysb.2012.07.028}{\textit{Nuclear Phys.~B}} \textbf{865}
 (2012), 133--147, \href{http://arxiv.org/abs/1203.4965}{arXiv:1203.4965}.

\bibitem{expansioin6}
Gurau R., The {$1/N$} expansion of tensor models beyond perturbation theory,
 \href{http://dx.doi.org/10.1007/s00220-014-1907-2}{\textit{Comm. Math. Phys.}} \textbf{330} (2014), 973--1019,
 \href{http://arxiv.org/abs/1304.2666}{arXiv:1304.2666}.

\bibitem{universality}
Gurau R., Universality for random tensors, \href{http://dx.doi.org/10.1214/13-AIHP567}{\textit{Ann. Inst. Henri Poincar\'e
 Probab. Stat.}} \textbf{50} (2014), 1474--1525, \href{http://arxiv.org/abs/1111.0519}{arXiv:1111.0519}.

\bibitem{RTM}
Gurau R., Random tensors, Oxford University Press, Oxford, 2016.

\bibitem{GW1}
Gurau R., Magnen J., Rivasseau V., Vignes-Tourneret F., Renormalization of
 non-commutative {$\Phi^4_4$} f\/ield theory in {$x$} space, \href{http://dx.doi.org/10.1007/s00220-006-0055-8}{\textit{Comm. Math.
 Phys.}} \textbf{267} (2006), 515--542, \href{http://arxiv.org/abs/hep-th/0512271}{hep-th/0512271}.

\bibitem{expansion2}
Gurau R., Rivasseau V., The $1/N$ expansion of colored tensor models in
 arbitrary dimension, \href{http://dx.doi.org/10.1209/0295-5075/95/50004}{\textit{Europhys. Lett.}} \textbf{95} (2011), 50004,
 5~pages, \href{http://arxiv.org/abs/1101.4182}{arXiv:1101.4182}.

\bibitem{LVE4}
Gurau R., Rivasseau V., The multiscale loop vertex expansion, \href{http://dx.doi.org/10.1007/s00023-014-0370-0}{\textit{Ann.
 Henri Poincar\'e}} \textbf{16} (2015), 1869--1897, \href{http://arxiv.org/abs/1312.7226}{arXiv:1312.7226}.

\bibitem{review}
Gurau R., Ryan J.P., Colored tensor models~-- a~review, \href{http://dx.doi.org/10.3842/SIGMA.2012.020}{\textit{SIGMA}}
 \textbf{8} (2012), 020, 78~pages, \href{http://arxiv.org/abs/1109.4812}{arXiv:1109.4812}.

\bibitem{melbp}
Gurau R., Ryan J.P., Melons are branched polymers, \href{http://dx.doi.org/10.1007/s00023-013-0291-3}{\textit{Ann. Henri
 Poincar\'e}} \textbf{15} (2014), 2085--2131, \href{http://arxiv.org/abs/1302.4386}{arXiv:1302.4386}.

\bibitem{GurSch}
Gurau R., Schaef\/fer G., Regular colored graphs of positive degree,
 \href{http://arxiv.org/abs/1307.5279}{arXiv:1307.5279}.

\bibitem{Gurau:2015tua}
Gurau R., Tanasa A., Youmans D.R., The double scaling limit of the
 multi-orientable tensor model, \href{http://dx.doi.org/10.1209/0295-5075/111/21002}{\textit{Europhys. Lett.}} \textbf{111} (2015),
 21002, 6~pages, \href{http://arxiv.org/abs/1505.00586}{arXiv:1505.00586}.

\bibitem{thomasmatrix}
Gurau R.G., Krajewski T., Analyticity results for the cumulants in a random
 matrix model, \href{http://dx.doi.org/10.4171/AIHPD/17}{\textit{Ann. Inst. Henri Poincar\'e~D}} \textbf{2} (2015),
 169--228, \href{http://arxiv.org/abs/1409.1705}{arXiv:1409.1705}.

\bibitem{Kazakov:1985ds}
Kazakov V.A., Bilocal regularization of models of random surfaces,
 \href{http://dx.doi.org/10.1016/0370-2693(85)91011-1}{\textit{Phys. Lett.~B}} \textbf{150} (1985), 282--284.

\bibitem{Kazakov:1986hu}
Kazakov V.A., Ising model on a dynamical planar random lattice: exact solution,
 \href{http://dx.doi.org/10.1016/0375-9601(86)90433-0}{\textit{Phys. Lett.~A}} \textbf{119} (1986), 140--144.

\bibitem{Kazakovmulticrit}
Kazakov V.A., The appearance of matter f\/ields from quantum f\/luctuations of
 {$2$}{D}-gravity, \href{http://dx.doi.org/10.1142/S0217732389002392}{\textit{Modern Phys. Lett.~A}} \textbf{4} (1989),
 2125--2139.

\bibitem{Kegeles:2016wfg}
Kegeles A., Oriti D., Continuous point symmetries in group f\/ield theories,
 \href{http://arxiv.org/abs/1608.00296}{arXiv:1608.00296}.

\bibitem{Knizhnik:1988ak}
Knizhnik V.G., Polyakov A.M., Zamolodchikov A.B., Fractal structure of
 {$2$}{D}-quantum gravity, \href{http://dx.doi.org/10.1142/S0217732388000982}{\textit{Modern Phys. Lett.~A}} \textbf{3} (1988),
 819--826.

\bibitem{sdequations2}
Krajewski T., Schwinger--{D}yson equations in group f\/ield theories of quantum
 gravity, in Symmetries and Groups in Contemporary Physics, \href{http://dx.doi.org/10.1142/9789814518550_0050}{\textit{Nankai
 Ser. Pure Appl. Math. Theoret. Phys.}}, Vol.~11, World Sci. Publ., Hackensack,
 NJ, 2013, 373--378, \href{http://arxiv.org/abs/1211.1244}{arXiv:1211.1244}.

\bibitem{Lahoche:2015tqa}
Lahoche V., Oriti D., Renormalization of a tensorial f\/ield theory on the
 homogeneous space {${\rm SU}(2)/{\rm U}(1)$}, \href{http://arxiv.org/abs/1506.08393}{arXiv:1506.08393}.

\bibitem{Lahoche:2016xiq}
Lahoche V., Samary D.O., Functional renormalisation group for the
 {$U(1)-T_5^6$} TGFT with closure constraint, \href{http://arxiv.org/abs/1608.00379}{arXiv:1608.00379}.

\bibitem{BrownMap1}
Le~Gall J.F., The topological structure of scaling limits of large planar maps,
 \href{http://dx.doi.org/10.1007/s00222-007-0059-9}{\textit{Invent. Math.}} \textbf{169} (2007), 621--670, \href{http://arxiv.org/abs/math.PR/0607567}{math.PR/0607567}.

\bibitem{BrownMap2}
Le~Gall J.F., Geodesics in large planar maps and in the {B}rownian map,
 \href{http://dx.doi.org/10.1007/s11511-010-0056-5}{\textit{Acta Math.}} \textbf{205} (2010), 287--360, \href{http://arxiv.org/abs/0804.3012}{arXiv:0804.3012}.

\bibitem{BrownMap3}
Le~Gall J.F., Uniqueness and universality of the {B}rownian map, \href{http://dx.doi.org/10.1214/12-AOP792}{\textit{Ann.
 Probab.}} \textbf{41} (2013), 2880--2960, \href{http://arxiv.org/abs/1105.4842}{arXiv:1105.4842}.

\bibitem{LVE3}
Magnen J., Rivasseau V., Constructive {$\phi^4$} f\/ield theory without tears,
 \href{http://dx.doi.org/10.1007/s00023-008-0360-1}{\textit{Ann. Henri Poincar\'e}} \textbf{9} (2008), 403--424,
 \href{http://arxiv.org/abs/0706.2457}{arXiv:0706.2457}.

\bibitem{Makeenko:1991ry}
Makeenko Yu., Loop equations and Virasoro constraints in matrix models,
 \href{http://arxiv.org/abs/hep-th/9112058}{hep-th/9112058}.

\bibitem{Eynard2}
Marchal O., Eynard B., Berg{\`e}re M., The sine-law gap probability,
 {P}ainlev\'e 5, and asymptotic expansion by the topological recursion,
 \href{http://dx.doi.org/10.1142/S2010326314500130}{\textit{Random Matrices Theory Appl.}} \textbf{3} (2014), 1450013, 41~pages,
 \href{http://arxiv.org/abs/1311.3217}{arXiv:1311.3217}.

\bibitem{Mehta}
Mehta M.L., Random matrices, \textit{Pure and Applied Mathematics (Amsterdam)},
 Vol.~142, 3rd~ed., Elsevier/Academic Press, Amsterdam, 2004.

\bibitem{oldgft6}
Ooguri H., Topological lattice models in four dimensions, \href{http://dx.doi.org/10.1142/S0217732392004171}{\textit{Modern Phys.
 Lett.~A}} \textbf{7} (1992), 2799--2810, \mbox{\href{http://arxiv.org/abs/hep-th/9205090}{hep-th/9205090}}.

\bibitem{Oriti:2011jm}
Oriti D., The microscopic dynamics of quantum space as a group f\/ield theory, in
 Foundations of Space and Time, Cambridge University Press, Cambridge, 2012,
 257--320, \href{http://arxiv.org/abs/1110.5606}{arXiv:1110.5606}.

\bibitem{Oriti:2014uga}
Oriti D., Group f\/ield theory and loop quantum gravity, \href{http://arxiv.org/abs/1408.7112}{arXiv:1408.7112}.

\bibitem{Oriti:2015qva}
Oriti D., Pranzetti D., Ryan J.P., Sindoni L., Generalized quantum gravity
 condensates for homogeneous geometries and cosmology, \href{http://dx.doi.org/10.1088/0264-9381/32/23/235016}{\textit{Classical
 Quantum Gravity}} \textbf{32} (2015), 235016, 40~pages, \href{http://arxiv.org/abs/1501.0093}{arXiv:1501.0093}.

\bibitem{Oriti:2015rwa}
Oriti D., Pranzetti D., Sindoni L., Horizon entropy from quantum gravity
 condensates, \href{http://dx.doi.org/10.1103/PhysRevLett.116.211301}{\textit{Phys. Rev. Lett.}} \textbf{116} (2016), 211301, 6~pages,
 \href{http://arxiv.org/abs/1510.06991}{arXiv:1510.06991}.

\bibitem{Oriti:2014yla}
Oriti D., Ryan J.P., Th{\"u}rigen J., Group f\/ield theories for all loop quantum
 gravity, \href{http://dx.doi.org/10.1088/1367-2630/17/2/023042}{\textit{New~J. Phys.}} \textbf{17} (2015), 023042, 46~pages,
 \href{http://arxiv.org/abs/1409.3150}{arXiv:1409.3150}.

\bibitem{Oriti:2016ueo}
Oriti D., Sindoni L., Wilson-Ewing E., Bouncing cosmologies from quantum
 gravity condensates, \href{http://arxiv.org/abs/1602.08271}{arXiv:1602.08271}.

\bibitem{Oriti:2016qtz}
Oriti D., Sindoni L., Wilson-Ewing E., Emergent {F}riedmann dynamics with a
 quantum bounce from quantum gravity condensates, \href{http://arxiv.org/abs/1602.05881}{arXiv:1602.05881}.

\bibitem{Politzer:1973fx}
Politzer H.D., Reliable perturbative results for strong interactions?,
 \href{http://dx.doi.org/10.1103/PhysRevLett.30.1346}{\textit{Phys. Rev. Lett.}} \textbf{30} (1973), 1346--1349.

\bibitem{LVE1}
Rivasseau V., Constructive matrix theory, \href{http://dx.doi.org/10.1088/1126-6708/2007/09/008}{\textit{J.~High Energy Phys.}}
 \textbf{2007} (2007), no.~9, 008, 13~pages, \href{http://arxiv.org/abs/0706.1224}{arXiv:0706.1224}.

\bibitem{LVE2}
Rivasseau V., Constructive f\/ield theory in zero dimension, \href{http://dx.doi.org/10.1155/2009/180159}{\textit{Adv. Math.
 Phys.}} \textbf{2009} (2009), 180159, 12~pages, \href{http://arxiv.org/abs/0906.3524}{arXiv:0906.3524}.

\bibitem{Rivasseau:2011hm}
Rivasseau V., Quantum gravity and renormalization: the tensor track,
 \href{http://dx.doi.org/10.1063/1.4715396}{\textit{AIP Conf. Proc.}} \textbf{1444} (2012), 18--29, \href{http://arxiv.org/abs/1112.5104}{arXiv:1112.5104}.

\bibitem{Rivasseau:2012yp}
Rivasseau V., The tensor track: an update, in Symmetries and Groups in
 Contemporary Physics, \href{http://dx.doi.org/10.1142/9789814518550_0011}{\textit{Nankai Ser. Pure Appl. Math. Theoret. Phys.}},
 Vol.~11, World Sci. Publ., Hackensack, NJ, 2013, 63--74, \href{http://arxiv.org/abs/1209.5284}{arXiv:1209.5284}.

\bibitem{Rivasseau:2014ima}
Rivasseau V., The tensor theory space, \href{http://dx.doi.org/10.1002/prop.201400057}{\textit{Fortschr. Phys.}} \textbf{62}
 (2014), 835--840, \href{http://arxiv.org/abs/1407.0284}{arXiv:1407.0284}.

\bibitem{Rivasseau:2013uca}
Rivasseau V., The tensor track,~{III}, \href{http://dx.doi.org/10.1002/prop.201300032}{\textit{Fortschr. Phys.}} \textbf{62}
 (2014), 81--107, \href{http://arxiv.org/abs/1311.1461}{arXiv:1311.1461}.

\bibitem{Ryan:2011qm}
Ryan J.P., Tensor models and embedded Riemann surfaces, \href{http://dx.doi.org/10.1103/PhysRevD.85.024010}{\textit{Phys. Rev.~D}}
 \textbf{85} (2012), 024010, 9~pages, \href{http://arxiv.org/abs/1104.5471}{arXiv:1104.5471}.

\bibitem{Salam1}
Salam A., Weak and electromagnetic interactions, in Elementary Particle Theory,
 Editor N.~Svartholm, Wiley, New York, Almqvist and Wiksell, Stockholm, 1968,
 367--377.

\bibitem{Samary:2013xla}
Samary D.O., Beta functions of ${\rm U}(1)^d$ gauge invariant just
 renormalizable tensor models, \href{http://dx.doi.org/10.1103/PhysRevD.88.105003}{\textit{Phys. Rev.~D}} \textbf{88} (2013),
 105003, 15~pages, \href{http://arxiv.org/abs/1303.7256}{arXiv:1303.7256}.

\bibitem{Samary:2014tja}
Samary D.O., Closed equations of the two-point functions for tensorial group
 f\/ield theory, \href{http://dx.doi.org/10.1088/0264-9381/31/18/185005}{\textit{Classical Quantum Gravity}} \textbf{31} (2014), 185005,
 29~pages, \href{http://arxiv.org/abs/1401.2096}{arXiv:1401.2096}.

\bibitem{Samary:2012bw}
Samary D.O., Vignes-Tourneret F., Just renormalizable {TGFT}'s on {${\rm
 U}(1)^d$} with gauge invariance, \href{http://dx.doi.org/10.1007/s00220-014-1930-3}{\textit{Comm. Math. Phys.}} \textbf{329}
 (2014), 545--578, \href{http://arxiv.org/abs/1211.2618}{arXiv:1211.2618}.

\bibitem{sasa1}
Sasakura N., Tensor model for gravity and orientability of manifold,
 \href{http://dx.doi.org/10.1142/S0217732391003055}{\textit{Modern Phys. Lett.~A}} \textbf{6} (1991), 2613--2623.

\bibitem{sasac}
Sasakura N., Super tensor models, super fuzzy spaces and super $n$-ary
 transformations, \href{http://dx.doi.org/10.1142/S0217751X11054449}{\textit{Internat.~J. Modern Phys.~A}} \textbf{26} (2011),
 4203--4216, \href{http://arxiv.org/abs/1106.0379}{arXiv:1106.0379}.

\bibitem{sasab}
Sasakura N., Tensor models and hierarchy of {$n$}-ary algebras,
 \href{http://dx.doi.org/10.1142/S0217751X1105381X}{\textit{Internat.~J. Modern Phys.~A}} \textbf{26} (2011), 3249--3258,
 \href{http://arxiv.org/abs/1104.5312}{arXiv:1104.5312}.

\bibitem{Gilles1}
Schaef\/fer G., Bijective census and random generation of {E}ulerian planar maps
 with prescribed vertex degrees, \textit{Electron.~J. Combin.} \textbf{4}
 (1997), 20, 14~pages.

\bibitem{Sindoni:2014wya}
Sindoni L., Ef\/fective equations for {GFT} condensates from f\/idelity,
 \href{http://arxiv.org/abs/1408.3095}{arXiv:1408.3095}.

\bibitem{'tHooft:1973jz}
't~Hooft G., A planar diagram theory for strong interactions, \href{http://dx.doi.org/10.1016/0550-3213(74)90154-0}{\textit{Nuclear
 Phys.~B}} \textbf{72} (1974), 461--473.

\bibitem{'tHooft:1972fi}
't~Hooft G., Veltman M., Regularization and renormalization of gauge f\/ields,
 \href{http://dx.doi.org/10.1016/0550-3213(72)90279-9}{\textit{Nuclear Phys.~B}} \textbf{44} (1972), 189--213.

\bibitem{tHooftgrav}
't~Hooft G., Veltman M., One-loop divergencies in the theory of gravitation,
 \textit{Ann. Inst. H.~Poincar\'e Sect.~A} \textbf{20} (1974), 69--94.

\bibitem{Tanasa:2011ur}
Tanasa A., Multi-orientable group f\/ield theory, \href{http://dx.doi.org/10.1088/1751-8113/45/16/165401}{\textit{J.~Phys.~A: Math.
 Theor.}} \textbf{45} (2012), 165401, 19~pages, \href{http://arxiv.org/abs/1109.0694}{arXiv:1109.0694}.

\bibitem{Weinberg1}
Weinberg S., A model of leptons, \href{http://dx.doi.org/10.1103/PhysRevLett.19.1264}{\textit{Phys. Rev. Lett.}} \textbf{19} (1967),
 1264--1266.

\bibitem{wigner}
Wigner E.P., Characteristic vectors of bordered matrices with inf\/inite
 dimensions, \href{http://dx.doi.org/10.2307/1970079}{\textit{Ann. of Math.}} \textbf{62} (1955), 548--564.

\bibitem{wishart}
Wishart J., The generalised product moment distribution in samples from
 a~normal multivariate population, \href{http://dx.doi.org/10.1093/biomet/20A.1-2.32}{\textit{Biometrika}} \textbf{20A} (1928),
 32--52.

\end{thebibliography}
\end{document}